\documentclass[%
 aip,
 jcp,%
 amsmath,amssymb,
 reprint,%
]{revtex4-1}

\newtheorem{lemm}{Lemma}
\usepackage{subfigure}
\usepackage{graphicx}
\usepackage{dcolumn}
\usepackage{bm}

\begin{document}

\preprint{AIP/123-QED}

\title{Efficiency analysis of diffusion on T-fractals in the sense of random walks}

\author{Junhao Peng}
\email{pengjh@gzhu.edu.cn}
\affiliation {College of Math and Information Science, Guangzhou University, Guangzhou 510006, China.}

\affiliation {Key Laboratory of Mathematics and Interdisciplinary Sciences of Guangdong
Higher Education Institutes, Guangzhou University, Guangzhou 510006, China}
\author{ Guoai Xu}
\affiliation {State Key Laboratory of Networking and Switching Technology, Beijing University of Posts and Telecommunications, Beijing 100876, China}

\date{\today}

\begin{abstract}
Efficiently controlling the diffusion  process is crucial in the study of diffusion problem in complex systems. In the sense of random walks with a single trap,  mean trapping time(MTT) and mean diffusing time(MDT) are good measures of  trapping efficiency and diffusion efficiency respectively. They both vary with the location of the node. In this paper, we study random walks on T-fractal and  provided general methods to calculate the MTT for any target node and the MDT  for any source  node. Using the MTT and the MDT as the measure of trapping efficiency and diffusion efficiency respectively, we compare the trapping efficiency and diffusion efficiency among  all nodes  of T-fractal and find the best (or worst)  trapping sites  and the best (or worst) diffusing sites. Our results show that: the hub node of T-fractal is the best trapping site, but it is also the worst diffusing site, the three boundary  nodes  are the worst trapping sites, but they are also the best diffusing sites. Comparing the minimum and maximum of MTT and MDT, we found that the maximum of MTT is almost $6$ times of the minimum  for MTT and the maximum of MDT is almost equal to the minimum  for MDT. These results show that  the location of target node has big effect on the trapping efficiency, but the location of source node almost has no effect on diffusion efficiency.
We also conducted numerical simulation to test the results we have  derived, the results we derived are consistent with those obtained by numerical simulation.
\end{abstract}


\maketitle


\section{Introduction}
\label{intro}
Efficiently controlling the diffusion  and transport process is crucial in the study of diffusion  and transport problem in complex systems. Random walks, which can be applied  as a model for diffusion and transport phenomena in complex systems, has given rise to a lot of interest in the past few years\cite{HaBe87, HaWe86, KaRe89, Mari89, BaKl98JCP, MariSaSt93, RaTo83, BeTuKo10}. Many problems in physics and chemistry are related to random walks on disordered media\cite{Ko00, Ki58, Avraham_Havlin04, BlZu81}.

In the sense of random walks with a single trap, a basic quantity relevant to random walks is the trapping time or mean first-passage time (MFPT), which is the expected number of steps to hit the target node(or trap) for the first time, for a walker starting from a source node. It is a quantitative indicator to characterize the transport efficiency and  many other quantities can be expressed in terms of it. Locating the target node(or trap) at one special node and average the MFPTs over all the source nodes, we get mean trapping time(MTT) for the special node. Locating the source node at one special node and  average the MFPTs over all the target nodes,  we obtain mean diffusing time(MDT) for the special node. Both the MTT and MDT  vary with the location of node and they are good measures of trapping efficiency and diffusion efficiency  respectively. Comparing the  MTT and  MDT  among  all the network nodes, we can find the effects of target node location on the trapping efficiency and the effects of source node location on diffusion efficiency. The nodes which  have the minimum MTT (or the maximum MTT) are best (or worst)  trapping sites and  the nodes which have the minimum MDT (or maximum MDT) are the best (or worst) diffusion sites.

 Because  the fractal structures are able to mimic a wide range of systems\cite{SongHaMa06, RoHa07, DoMe03, AlBa02}, in the past several years, random walks on fractals, especially the MFPT on different deterministic fractals, has been extensively studied\cite{BeTuKo10, Mo69, GiMaNa94, KoBa02, MoHa89, BeMeTe08, HaRo08, FuDoBl13, HeMaKn04}. For example, the MTT for some special nodes were  obtained for different  fractals(or networks), such as   Sierpinski  gaskets\cite{KoBa02}, Apollonian network\cite{ZhGuXi09}, pseudofractal scale-free web \cite{ZhQiZh09}, deterministic scale-free graph\cite{AgBuMa10} and some special trees\cite{CoMi10,  ZhZhGa10, LiZh13, LiWuZh11, WuLiZhCh12, ZhLi11}.  The  global mean first-passage time (i.e. the average of MFPTs over all pairs of nodes) were  obtain for some special trees \cite{ZhZhGa10, LiZh13, LiWuZh11, ZhLi11, ZhWu10} and dual Sierpinski gaskets\cite{WuZh11}. The MDT were obtained for  exponential treelike networks\cite{ZhLiLin11},  scale-free Koch networks\cite{ZhGa11} and deterministic scale-free graph\cite{AgBu09}.

 However, the results of MTT and MDT which were obtained are only restricted to some special nodes for the above networks and we can neither compare   the MTTs(or MDTs)  among all the network nodes nor  analyze the effect of nodes location on the trapping efficiency and diffusion efficiency.  It is still difficult to deriving the analytic solutions of  the MTT for any target node  and the MDT for any source node. It is also difficult to deriving the analytic solutions of MFPT for any pair of nodes.

   As for T-fractal, it is a special tree, the MTT for the hub node and the GMFPT had been obtained\cite{MeAgBeVo12, Agl08, ZhYu09}. The MTT for some low-generation nodes can also be derived due to the methods of Ref. \cite{MeAgBeVo12}. But the  analytic calculations of MFPT for any pair of nodes, the MTT for any target node and the MDT for any source node were still unresolved.

 In this paper, we study random walks  on T-fractal  based on its self-similar structure and the relations between  random walks and electrical networks\cite{Te91, LO93}.  We first provided general methods for calculating the MFPT between any pair of nodes, the MTT for any target node and MDT  for any source  node,  then calculated MFPT, the MTT and MDT for some special nodes.  We also conducted numerical simulation to test the results we have  derived, the results we derived are consistent with those obtained by numerical simulation.   Further more, using the MTT and the MDT as the measures of trapping efficiency and diffusion efficiency respectively, we compare the trapping efficiency and diffusion efficiency among all the nodes  of T-fractal and find the best (or worst)  trapping sites and the best (or worst) diffusing sites. Our results show that: the hub node of T-fractal is the best trapping site, but it is also the worst diffusing site,  the three boundary  nodes are the worst trapping sites, but they are also the best diffusing sites.

  Comparing the minimum and maximum for MTT and MDT, we found that the maximum of MTT is almost $6$ times of the minimum  for MTT and the maximum of MDT is almost equal to the minimum  of MDT. These results show that  the trap's position has large effect on the trapping efficiency, but the position of source node has little effect on diffusion efficiency.
  The methods  we presented can also be used to solve the problems of MFPT on other self-similar trees.

\section{The network model}
\label{sec:1}
\begin{figure}
\begin{center}
\includegraphics[scale=0.5]{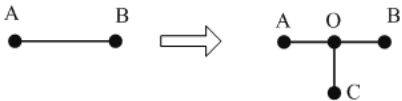}
\caption{Iterative construction method of T-fractal}
\label{Edge_replace}       
\end{center}
\end{figure}

 Here, The T-fractal we considered is constructed iteratively\cite{Agl08}. For convenience, we call the  times of iterations as the generation of the T-fractal and  denote by $G(t)$ the T-fractal of generation $t$. For $t = 0, G(0)$ is an edge connecting two nodes. For $t >0, G (t)$ is obtained from $G (t-1)$ via replacing every edge in $G(t-1)$ by a ``T'' structure illustrated in FIG. \ref{Edge_replace}.

\begin{figure}
\begin{center}
\includegraphics[scale=0.5]{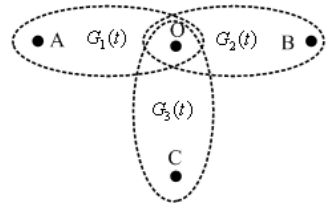}
\caption{Alternative construction of the T-fractal which highlights self-similarity: the T-fractal of generation $t$, denoted by $G(t)$, is composed of three copies of $G(t-1)$ labeled as  $G_1(t)$,$G_2(t)$,$G_3(t)$
}
\label{structure}
\end{center}
\end{figure}

The T-fractal $G(t)$ can also be constructed by another method which is shown in FIG.\ref{structure}: the T-fractal $G(t)$  is composed of $3$ copies, called subunit, of $G(t-1)$ which are  joined  at the hub node O(i.e., nodes at the center of $G(t)$ ).  According to its construction, the total number of edges for $G(t)$ is $E_t=3^t$ and the total number of nodes for $G(t)$ satisfies\cite{Agl08}
 \begin{equation}
N_t=1+E_t=1+3^t
\label{eq_nodes}
\end{equation}

\section{Formulation of the problem}
\label{sec:gen_meth}
In this paper, we study unbiased discrete random walks  on the T-fractal, at each time step, the particle
(walker), starting from its current location, moves to any of its nearest neighbors with equal probability.
 The quantity we are interested in is  mean first-passage time (MFPT), which is the expected number of steps to hit the target node(or trap) for the first time, for a walker starting from a source node.

Let $F(x,y)$  denote the MFPT from nodes $x$ to $y$ in T-fractal $G(t)$, The sum
$$k(x,y)=F(x,y)+F(x,y)$$ is called the commute time and the MFPT can be expressed in term  of commute times\cite{Te91}.
\begin{equation}
F(x,y)=\frac{1}{2}\!\left\{\!k(x,y)\!+\!\!\sum_{u\in G(t)}\!\pi(u)[k(y,u)-k(x,u)]\!\right\}\!
\label{FXY}
\end{equation}
where \lq\lq$u\in G(t)$\rq\rq means that $u$ belongs to the nodes set of $G(t)$ , $\pi(u)=\frac{d_u}{2E_t}$ is the stationary distribution for random walks on the T-fractal  and $d_u$ is the degree of node $u$.

If we view the networks under consideration as electrical networks  by considering each edge to be a unit resistor and let $\Psi_{xy}$ denote the effective resistance  between two nodes $x$ and $y$ in the electrical networks,
 we have\cite{Te91}
\begin{equation}
k(x,y)=2E_t\Psi_{xy}
\label{KR}
\end{equation}\par
 where $E_t$ is the total numbers of edges of $G(t)$. Since the T-fractals we studied  are trees, the effective resistance between any two nodes is exactly the  shortest-path length  between the two nodes. Hence
 \begin{equation}
 \Psi_{xy}=L_{xy}
 \end{equation}
 where $L_{xy}$ denote  the shortest path length between node $x$ to node $y$.  Thus
 \begin{equation}
k(x,y)=2E_tL_{xy}
\label{KL}
\end{equation}
Inserting Eq.(\ref{KL}) for $k(x,y)$ into Eq.(\ref{FXY}),we obtain
\begin{eqnarray}
F(x,y)&=&E_t\!\left\{\!L_{xy}\!+\!\!\sum_{u\in G(t)}\!\pi(u)L_{yu}\!-\!\!\sum_{u\in G(t)}\!\pi(u)L_{xu} \!\right\}\!
\label{FXYL}
\end{eqnarray}

If we average the MFPTs over all the source nodes and all target nodes, we  obtain MTT and MDT respectively. That is to say, if we   define
\begin{eqnarray}
T_y&=&\frac{1}{E_t}\sum_{x\in G(t),x\neq y}F(x,y) \label{MTTo}\\
D_x&=&\frac{1}{E_t}\sum_{y\in G(t),y\neq x}F(x,y)  \label{MSTo}
\end{eqnarray}
$T_y$ is just the mean trapping time(MTT) for target node $y$ and $D_x$ is just mean diffusing time(MDT) for source node $x$.
Let
\begin{equation}
S_x= \sum_{y \in G(t)}{L_{xy}}
\label{SX}
\end{equation}
\begin{equation}
W_x=\sum_{u\in G(t)}\pi(u)L_{xu}=\frac{1}{2E_t}\cdot \sum_{u \in G(t)}{(L_{xu}\cdot d_u)}
\label{WY}
\end{equation}
\begin{equation}
\Sigma=\sum_{u\in G(t)}\left(\pi(u)\sum_{x\in G(t)}L_{xu}\right)
\label{WS}
\end{equation}
 and substitute  $F(x,y)$ with Eq.(\ref{FXYL}) in Eqs.(\ref{MTTo}) and (\ref{MSTo}), we obtain
\begin{eqnarray} \label{MTT}
T_y
&=&S_y+N_t\cdot W_y-\Sigma
\end{eqnarray}
\begin{eqnarray} \label{MDT}
D_x
   &=&S_x+\Sigma-N_t\cdot W_x
\end{eqnarray}
Hence, if we can calculate $\Sigma$ and $S_x, W_x$ for any node $x$, we can calculate $F(x,y)$ for any two nodes $(x,y)$ and  MTT and MDT for any  node $x$. Although it is difficult to calculate these quantities for general tree, we presented methods for calculating these quantities for T-fractal based on its self-similar structure. Therefore, we can calculate  MTT and MDT for any  node.
\section{The methods for calculating MTT and MDT}
\label{sec:det_meth}
 \subsection{Methods for calculating $S_x$ and $W_x$}
 \label{Met_SW}
\begin{figure}
\begin{center}
\includegraphics[scale=0.3]{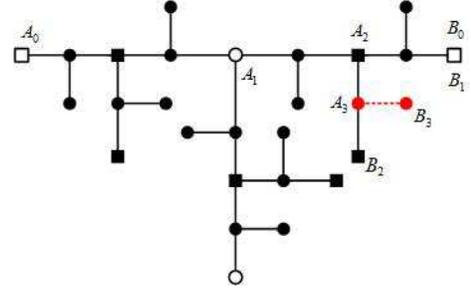}
\caption{The T-fractal of generation $3$: levels. Hollow square, level 0,1,2,3; hollow circle, level 1,2,3; solid square, level 2,3; solid  circle, level 3.}
\label{label_subunit}
\end{center}
\end{figure}

 For convenience, we classify the nodes of $G(t)$ into different  levels. Nodes, which are introduced into the network before $k$(include $k$)  times of iterations, are said to belong to level $k$ in this paper. Thus nodes which belong to level $k$ also belong to level $k+1$, $k+2$, $\cdots, t$. For example, in T-fractal of generation $3$ which is shown in  FIG. \ref{label_subunit}, nodes $A_0,B_0$, which are  represented by hollow square, belong to level $0$. They are also belong to level $1, 2, 3$. Nodes represented by hollow circle belong to level $1, 2, 3$. Nodes represented by solid square belong to level $2, 3$. Nodes represented by solid circle belong to level $3$.

As shown in FIG. \ref{structure}, The T-fractal $G(t)$  is composed of $3$  subunits which are copies  of $G(t-1)$ and $G(t-1)$ is also composed of $2m+2$ subunits which are copies  of $G(t-2)$ . In order to tell apart the different  structures of these subunits, we  classify these subunits  into different levels and let $\Lambda_k$ denote the subunit of level $k(k\geq0)$. In this paper, $G(t)$ is said to be subunit of level $0$. For any   $k\geq0$, the $3$  subunits of $\Lambda_k$ are said to be  subunits of level $k+1$. Thus, any edge of $G(t)$  is a subunit of level $t$ and $\Lambda_k$ is a copy of  T-fractal with generation $t-k$.

\begin{figure}
\begin{center}
\includegraphics[scale=0.3]{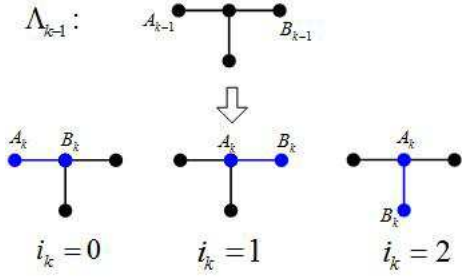}
\caption{Subunit represented by blue line are the  subunit $\Lambda_{k}$ corresponding to value of $i_k$ below, its two boundary nodes are labeled as $A_k, B_k$. Nodes labeled as $A_{k-1}$ in $\Lambda_{k-1}$ is also labeled as $A_k$ in $\Lambda_{k}$ while $i_k=0$. Nodes labeled as $B_{k-1}$ in $\Lambda_{k-1}$ is also labeled as $B_k$ in $\Lambda_{k}$ while $i_k=1$.}
\label{subn_k}       
\end{center}
\end{figure}

 In order to distinguish the subunits of different locations, similar to the method of Ref\cite{MeAgBeVo12}, we label the subunit  $\Lambda_k(1\leq k \leq t)$ by a sequence $\{i_1, i_2, ..., i_{k} \}$, where $i_j((1\leq j \leq k))$ labels its position in its father subunit $\Lambda_{j-1}$. FIG. \ref{subn_k} shows the construction of $\Lambda_{k-1}$ and the relation  between the value of $i_k$ and the location of  subunit $\Lambda_{k}$ in $\Lambda_{k-1}$: all subunit $\Lambda_{k}$ are represented by an edge, the one represented by blue edge are the  subunit $\Lambda_{k}$ corresponding to value of $i_k=0, 1, 2$.

 For example, in the  T-fractal of generation $3$ shown in  FIG. \ref{label_subunit}, the subunit  represented by red dotted line,   which is a subunit of level $3$,  is labeled by a sequence $\{1, 2, 2 \}$.\par
  For convenience, we label the two boundary nodes of subunit $\Lambda_k$ as $A_k, B_k$  and building mapping between  boundary nodes of $\Lambda_{k-1}$ and those of $\Lambda_{k}$ as shown in FIG. \ref{subn_k}.  The boundary node of $\Lambda_{k-1}$ labeled as $A_{k-1}$ is also a boundary node of $\Lambda_{k}$ labeled as $A_k$  while $i_k=0$. The boundary node of $\Lambda_{k-1}$ labeled as $B_{k-1}$ is also a boundary node of $\Lambda_{k}$ labeled as $B_k$ while $i_k=1$.

 Let
\begin{equation}\label{Def_SK}
  \mathcal{S}^{(k)}\equiv \left( \begin{array}{c} S_{A_k}\\S_{B_k} \end{array} \right)
\end{equation}
\begin{equation}\label{Def_WK}
  \mathcal{W}^{(k)}\equiv \left( \begin{array}{c} W_{A_k}\\W_{B_k} \end{array} \right)
\end{equation}

As derived in Appendix \ref{Pof_Rec_sk} and Appendix \ref{Pof_Rec_wk}, we obtain the following results.

\begin{lemm} \label{Rec_Sk}  
 For any  $k>0$ , $\mathcal{S}^{(k)}$  satisfy the following recursion relations
\begin{equation}\label{SK}
  \hspace{-10mm}\mathcal{S}^{(k)}=\mathcal{M}_{i_k}\mathcal{S}^{(k-1)}+\mathcal{V}_{i_k}^k, \quad  i_k=0, 1, 2
\end{equation}
where
\begin{equation} \label{MV0}
\mathcal{M}_{0}=
\left(                 
  \begin{array}{cc}
   1 & 0 \\
   1/2 & 1/2
  \end{array}
\right),
 \mathcal{V}_{0}^k=\left( \begin{array}{c} 0 \\ -2\cdot6^{t-k} \end{array} \right)                
 \end{equation}
\begin{equation} \label{MV1}
\mathcal{M}_{1}=
\left(                 
  \begin{array}{cc}
   1/2 & 1/2 \\
   0 & 1
  \end{array}
\right),                 
\mathcal{V}_{1}^k=\left( \begin{array}{c} -2\cdot6^{t-k}\\0 \end{array} \right) 
\end{equation}
\begin{equation} \label{MV2}
\mathcal{M}_{2}=
\left(                 
  \begin{array}{cc}
   1/2 & 1/2 \\
   1/2 & 1/2
  \end{array}
\right),
\mathcal{V}_{2}^k=\left( \begin{array}{c} -2\cdot6^{t-k}\\(3^t-3^{t-k+1})2^{t-k} \end{array} \right)
\end{equation}
\end{lemm}

\begin{lemm} \label{Rec_Wk}  
 For any  $k>0$ , $\mathcal{W}^{(k)}$  satisfy the following recursion relations
\begin{equation}\label{WK}
  \mathcal{W}^{(k)}=\mathcal{M}_{i_k}\mathcal{W}^{(k-1)}+\mathcal{U}_{i_k}^k, \quad  i_k=0, 1, 2
\end{equation}
where $\mathcal{M}_{i_k}(i_k=0, 1, 2)$ are given by Eqs.(\ref{MV0}), (\ref{MV1}), (\ref{MV2}) and  $\mathcal{U}_{i_k}^k(i_k=0, 1, 2)$  are given by
\begin{equation} \label{U01}
\mathcal{U}_{0}^k=\left( \begin{array}{c} 0 \\ -\frac{2^{t-k+1}}{3^k} \end{array} \right),                  
\mathcal{U}_{1}^k=\left( \begin{array}{c} -\frac{2^{t-k+1}}{3^k}\\0 \end{array} \right)                 
\end{equation}
\begin{equation} \label{U2}
\mathcal{U}_{2}^k=\left( \begin{array}{c} -\frac{2^{t-k+1}}{3^k}\\2^{t-k}(1-\frac{1}{3^{k-1}}) \end{array} \right)
\end{equation}
\end{lemm}

Using equation (\ref{SK}) repeatedly, we obtain
\begin{eqnarray}\label{SK0}
  \mathcal{S}^{(t)}&=&\mathcal{M}_{i_t}\mathcal{S}^{(t-1)}+\mathcal{V}_{i_t}^t  \nonumber \\
   &=&\mathcal{M}_{i_t}[\mathcal{M}_{i_{t-1}}\mathcal{S}^{(t-2)}+\mathcal{V}_{i_{t-1}}^{t-1}]+\mathcal{V}_{i_t}^t \nonumber \\
   &=& \mathcal{M}_{i_t}\mathcal{M}_{i_{t-1}}\mathcal{S}^{(t-2)}+\mathcal{M}_{i_t}\mathcal{V}_{i_{t-1}}^{t-1}+\mathcal{V}_{i_t}^t \nonumber \\
   &=& \cdots \nonumber \\
   &=&\mathcal{M}_{i_t}\mathcal{M}_{i_{t-1}}\cdots \mathcal{M}_{i_1}\mathcal{S}^{(0)}\nonumber \\
   & &+\sum_{l=1}^{t-1}\mathcal{M}_{i_t}\mathcal{M}_{i_{t-1}}\cdots \mathcal{M}_{i_{l+1}}\mathcal{V}_{i_{l}}^{l}+\mathcal{V}_{i_t}^t
\end{eqnarray}

Similarity
\begin{eqnarray}\label{WK0}
  \mathcal{W}^{(k)}
   &=&\mathcal{M}_{i_k}\mathcal{M}_{i_{k-1}}\cdots \mathcal{M}_{i_1}\mathcal{W}^{(0)}\nonumber \\
   &&+\sum_{l=1}^{k-1}\mathcal{M}_{i_k}\mathcal{M}_{i_{k-1}}\cdots \mathcal{M}_{i_{l+1}}\mathcal{U}_{i_{l}}^{l}+\mathcal{U}_{i_k}^k
\end{eqnarray}

 As for  $\mathcal{S}^{(0)}$ and $\mathcal{W}^{(0)}$, it is easy to know
 \begin{eqnarray}
 \mathcal{S}^{(0)}=(S_{A_0} \quad S_{B_0})^T=S_{A_0}(1 \quad 1)^T  \label{S0} \\
  \mathcal{W}^{(0)}=(W_{A_0} \quad W_{B_0})^T=W_{A_0}(1 \quad 1)^T \label{W0}
  \end{eqnarray}
  where $(x \quad y)^T$ is the transpose of vector  $(x \quad y)$  and $S_{A_0}$ and $W_{A_0}$ are $S_x$ and $T_x$ for nodes of level $0$ respectively, which have  been derived in Appendix \ref{Der_SW0}.

 Noticing that any edge of $G(t)$ is a subunit of level $t$, its two end nodes are just its two boundary nodes.  If we know its location (i.e. the label  sequence $\{i_1, i_2, ..., i_{t} \}$) for any edge of $G(t)$, we can exactly calculate $\mathcal{S}^{(t)}$  and $\mathcal{W}^{(t)}$ for its two end nodes. Hence, we can derive the expression of $S_x$  and $W_x$ for any node $x$ of $G(t)$.

\subsection{Exact calculation of $\Sigma$ }
\label{SSWP}

We find that
\begin{equation} \label{xigma}
  \Sigma=\sum_{u\in  G(t)}(\pi(u)\sum_{x\in  G(t)}L_{xu})=\frac{1}{2E_t}\sum_{u\in  G(t)}(d_uS_u)
 \end{equation}
$\sum_{u\in \Omega}(d_uS_u)$ is just the summation of  $S_x$ for the two end nodes of every edges of $G(t)$(Note: for node $x$ which is the intersection of $n$ edges,  $S_x$  will be counted $n$ times).   Because any edge of $G(t)$ is a subunit of level $t$, which is in one to one correspondence with a sequence  $\{i_1,\cdots,i_t\}$, its two end nodes are also its two boundary nodes labeled as $A_t, B_t$.  Thus
\begin{equation} \label{Uxigma}
  \sum_{u\in  G(t)}(d_uS_u)=\sum\left(\sum_{\{i_1,\cdots,i_t\}}{\mathcal{S}^{(t)}}\right)
\end{equation}
for the right side of the equation, the second summation is run over all the subunits of  level $t$(i.e. let $\{i_1,\cdots,i_t\}$ run over all the possible values), the first summation is just add the two entries of $\sum_{\{i_1,\cdots,i_t\}} \mathcal{S}^{(t)}$ together.

Let
\begin{equation} \label{MTotal}
\mathcal{M}_{tot} = \mathcal{M}_{0}+\mathcal{M}_{1}+ \mathcal{M}_{2}
\end{equation}
\begin{equation} \label{VTotal}
\mathcal{V}_{tot}^l = \mathcal{V}_{0}^{l}+\mathcal{V}_{1}^{l} +\mathcal{V}_{2}^{l}
\end{equation}
 we have
\begin{eqnarray}\label{SuoV}
  \sum_{\{i_1,\cdots,i_t\}}\mathcal{M}_{i_t}\mathcal{M}_{i_{t-1}}\cdots \mathcal{M}_{i_{l+1}}\mathcal{V}_{i_{l}}^{l}= 3^{l-1}\mathcal{M}_{tot}^{t-l}\mathcal{V}_{tot}^l
\end{eqnarray}
Thus
\begin{eqnarray}\label{SST}
  \sum_{\{i_1,\cdots,i_t\}} \mathcal{S}^{(t)}&=& \sum_{\{i_1,\cdots,i_t\}}( \mathcal{M}_{i_t}\mathcal{M}_{i_{t-1}}\cdots \mathcal{M}_{i_1}\mathcal{S}^{(0)}  \nonumber \\
    &&+\sum_{l=1}^{t-1}\mathcal{M}_{i_t}\mathcal{M}_{i_{t-1}}\cdots \mathcal{M}_{i_{l+1}}\mathcal{V}_{i_{l}}^{l}+\mathcal{V}_{i_t}^t ) \nonumber \\
   &=& \mathcal{M}_{tot}^{t}\mathcal{S}^{(0)}+\sum_{l=1}^{t}3^{l-1}\mathcal{M}_{tot}^{t-l}\mathcal{V}_{tot}^l
\end{eqnarray}

 Substituting  $\mathcal{M}_{i}$ with Eqs.(\ref{MV0}), (\ref{MV1}) and (\ref{MV2}) in Eq. (\ref{MTotal}), and orthogonal decomposing $\mathcal{M}_{tot}$, we obtain
 \begin{eqnarray} \label{MTD}
\mathcal{M}_{total}&=&
\left(                 
  \begin{array}{cc}
   2 & 1 \\
   1 & 2
  \end{array}
\right)\nonumber \\
&=&
\left(                 
  \begin{array}{cc}
   -\frac{\sqrt{2}}{2} & \frac{\sqrt{2}}{2} \\
   \frac{\sqrt{2}}{2} & \frac{\sqrt{2}}{2}
  \end{array}
\right)                 
\left(                 
  \begin{array}{cc}
   1 & 0 \\
   0 & 3
  \end{array}
\right)                 
\left(                 
  \begin{array}{cc}
   -\frac{\sqrt{2}}{2} & \frac{\sqrt{2}}{2} \\
   \frac{\sqrt{2}}{2} & \frac{\sqrt{2}}{2}
  \end{array}
\right)                 
\end{eqnarray}
Therefore
\begin{equation} \label{MTK}
\mathcal{M}_{total}^k=
\left(                 
  \begin{array}{cc}
   -\frac{\sqrt{2}}{2} & \frac{\sqrt{2}}{2} \\
   \frac{\sqrt{2}}{2} & \frac{\sqrt{2}}{2}
  \end{array}
\right)                 
\left(                 
  \begin{array}{cc}
   1 & 0 \\
   0 & {3^k}
  \end{array}
\right)                 
\left(                 
  \begin{array}{cc}
   -\frac{\sqrt{2}}{2} & \frac{\sqrt{2}}{2} \\
   \frac{\sqrt{2}}{2} & \frac{\sqrt{2}}{2}
  \end{array}
\right)                 
\end{equation}
 Similarity 
 \begin{equation} \label{VT}
\mathcal{V}_{tot}^l=\left( \begin{array}{c} -4 \\ 3^l-5 \end{array} \right)6^{t-l}
\end{equation}
 Thus
 \begin{eqnarray}\label{MTS0}
 \mathcal{M}_{tot}^{t}\mathcal{S}^{(0)}&=&
  \!\left(  \!               
  \begin{array}{cc}
   -\frac{\sqrt{2}}{2} & \frac{\sqrt{2}}{2} \\
   \frac{\sqrt{2}}{2} & \frac{\sqrt{2}}{2}
  \end{array}
\!\right)\!                 
\!\left( \!                
  \begin{array}{cc}
   1 & 0 \\
   0 & {3^t}
  \end{array}
\!\right) \!                
\!\left( \!                
  \begin{array}{cc}
   -\frac{\sqrt{2}}{2} & \frac{\sqrt{2}}{2} \\
   \frac{\sqrt{2}}{2} & \frac{\sqrt{2}}{2}
  \end{array}
\!\right) \!                
\!\left(\! \begin{array}{c} 1\\1\end{array} \!\right)\!S_{A_0}
 \nonumber \\
 &=& (4\cdot6^{t-1}+3^{t-1})3^t
     \!\left( \!\begin{array}{c} 1\\ 1 \end{array} \!\right)\!
\end{eqnarray}
and
\begin{eqnarray}\label{MTVT}
 &&\sum_{l=1}^{t}3^{l-1}\mathcal{M}_{tot}^{t-l}\mathcal{V}_{tot}^l\nonumber \\
&=&\!\sum_{l=1}^{t}\!3^{l-1}6^{t-l}
\!\left(\! \begin{array}{c}\frac{(3^l-5)}{2}3^{t-l}-2\cdot3^{t-l}-\frac{3^l-5}{2}-2 \\\frac{(3^l-5)}{2}3^{t-l}-2\cdot3^{t-l}+\frac{3^l-5}{2}+2\end{array}\!\right)\!
\end{eqnarray}

Inserting Eqs. (\ref{MTS0}), (\ref{MTVT}) and (\ref{SA}) into Eq.(\ref{SST}),
we obtain
\begin{eqnarray}\label{SSTC}
  &&\sum_{\{i_1,\cdots,i_t\}} \mathcal{S}^{(t)}
      =\mathcal{M}_{tot}^{t}\mathcal{S}^{(0)}+\sum_{l=1}^{t}3^{l-1}\mathcal{M}_{tot}^{t-l}\mathcal{V}_{tot}^l\nonumber \\
   &=&3^t\cdot6^{t}\left(\frac{8}{15}+\frac{1}{6\cdot2^t}+\frac{3}{10\cdot6^{t}}\right)\left( \begin{array}{c} 1\\1 \end{array} \right)
\end{eqnarray}

Replacing $\sum_{\{i_1,\cdots,i_t\}}\mathcal{S}^{(t)}$ with Eq.(\ref{SSTC}) in Eq. (\ref{Uxigma}),
we get
\begin{eqnarray}\label{SAPL}
 \Sigma
   &=&\frac{1}{2E_t}\sum_{u\in G(t)}\left(d_u S_u\right)\nonumber \\
   &=&6^{t}\left(\frac{8}{15}+\frac{1}{6\cdot2^t}+\frac{3}{10\cdot6^{t}}\right)
\end{eqnarray}

\subsection{Examples}
\label{sec:example}

According to the methods presented in Sec.(\ref{Met_SW})   and  Sec.(\ref{SSWP}),  we can calculate  $T_x$ and $D_x$ for any node $x$ of $G(t)$. 
 In order to explain our methods,   we calculate $S_x,  W_x$ for the three boundary nodes and the hub node labeled as $A, B, C, O$ respectilely(see FIG. \ref{structure}),   and then calculate the MFPT between these nodes, the  MTT and MDT for these nodes.

For nodes  $A$ and $B$, according to Eqs. (\ref{SA}) and (\ref{WA}), we obtain
\begin{equation}
S_A=S_B=4\cdot6^{t-1}+3^{t-1}\label{SAB}
\end{equation}
\begin{equation}
W_A=W_B=\frac{2^{t+1}}{3}-\frac{1}{6}\label{WAB}
\end{equation}

For nodes  $O$ and $C$, according to Eqs. (\ref{SOC}), (\ref{WO}) and (\ref{WC}),  we have
\begin{eqnarray}\label{SOCE}
  \left( \begin{array}{c} S_{O}\\S_{C} \end{array} \right)
  &=&\left(                 
  \begin{array}{cc}
   1/2 & 1/2 \\
   1/2 & 1/2
  \end{array}
\right)
 \left( \begin{array}{c}  S_{A}\\S_{B} \end{array} \right)
 +\left( \begin{array}{c} -2\cdot6^{t-k}\\(3^t-3^{t-k+1})\cdot2^{t-k} \end{array} \right) \nonumber  \\
  &=& \left( \begin{array}{c} 2\cdot6^{t-1}+3^{t-1}\\ 4\cdot6^{t-1}+3^{t-1} \end{array} \right)
\end{eqnarray}
and
\begin{eqnarray}\label{WOCE}
  \left( \begin{array}{c} W_{O}\\W_{C} \end{array} \right)
  &=&\left(                 
  \begin{array}{cc}
   1/2 & 1/2 \\
   1/2 & 1/2
  \end{array}
\right)
 \left( \begin{array}{c}  W_{A}\\W_{B} \end{array} \right)
 +\left( \begin{array}{c} -\frac{2^{t-k+1}}{3^k} \\(1-\frac{1}{3^{k-1}})\cdot 2^{t-k} \end{array} \right) \nonumber  \\
 &=& \left( \begin{array}{c} \frac{2^{t}}{3}- \frac{1}{6} \\ \frac{2^{t+1}}{3}- \frac{1}{6} \end{array} \right)
\end{eqnarray}

Thus, the MFPTs between any two nodes of  $A$, $B$, $O$  and $C$, which can be derived from Eq. (\ref{FXYL}), are as follows.
\begin{eqnarray}
F(A,C)&=&F(C,A)=F(C,B)=F(B,C)\nonumber\\
    &=&F(B,A)=F(A,B) \nonumber\\
      &=&E_t(L_{AB}+W_B-W_A) \nonumber\\
        &=&6^t \label{MFPTABC}
\end{eqnarray}
\begin{eqnarray}
F(O,C)&=&F(O,A)=F(O,B)\nonumber\\
      &=&E_t(L_{OB}+W_B-W_O) \nonumber\\
        &=&5\cdot6^{t-1} \label{MFPOABC}
\end{eqnarray}
\begin{eqnarray}
F(C,O)&=&F(B,O)=F(A,O)
        =6^{t-1} \label{MFPABCO}
\end{eqnarray}

Insert Eqs.(\ref {SAPL}), (\ref{SAB}), (\ref{WAB}), (\ref{SOCE})  and (\ref{WOCE}) into Eq.(\ref{MTT}), we obtain the MTT for node   $A$, $B$, $C$  and $O$.
\begin{eqnarray} \label{MTTABC}
T_A&=&T_B=T_C
=S_A+N_t W_A-\Sigma  \nonumber  \\
&=&6^t\cdot\left(\frac{4}{5}+\frac{2}{3^{t+1}}-\frac{14}{5\cdot6^{t+1}}\right)
\end{eqnarray}
\begin{eqnarray} \label{MTTO}
T_O
&=&6^{t-1}\cdot\left(\frac{4}{5}+\frac{2}{3^{t}}-\frac{14}{5\cdot6^{t}}\right)
\end{eqnarray}

These results  are consistent with those  derived in Ref. \cite{MeAgBeVo12,Agl08,ZhYu09}. 
Insert Eqs.(\ref {SAPL}), (\ref{SAB}), (\ref{WAB}), (\ref{SOCE})  and (\ref{WOCE})  into Eq.(\ref{MDT}), we obtain
the MDT for  node   $A$, $B$, $C$ and $O$.
\begin{eqnarray} \label{MDTABC}
D_A&=&D_B=D_C
   =S_A+\Sigma-N_t W_A  \nonumber  \\
   &=&
   6^t\cdot\left(\frac{8}{15}+\frac{1}{3\cdot2^{t-1}}+\frac{14}{5\cdot6^{t+1}}-\frac{2}{3^{t+1}}\right)
\end{eqnarray}
\begin{eqnarray} \label{MDTO}
D_O
   &=&6^{t}\cdot\left(\frac{8}{15}+\frac{1}{3\cdot 2^{t-1}}+\frac{14}{5\cdot6^{t+1}}-\frac{1}{3^{t+1}}\right)
\end{eqnarray}

We also conducted numerical simulation to test the results we have just derived, the results just derived are consistent with those obtained by numerical simulation.

\section{Comparison the trapping efficiency among all the nodes of T-fractal}
\label{sec:Com_MTT}

In these section, using the MTT  as the measure of trapping efficiency, we compare the trapping efficiency (i.e. the MTT) among all the nodes of T-fractal and find the best trapping sites(i.e. nodes which have the minimum MTT) and the  worst trapping sites(i.e. nodes which have the maximum MTT).

\begin{figure}
\begin{center}
\includegraphics[scale=0.2]{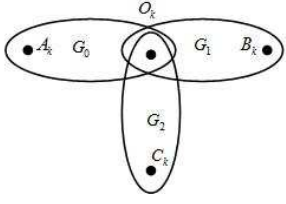}
\caption{Labels for the $4$ nodes of level $k+1$ in  $\Lambda_k$.}
\label{sub_k}       
\end{center}
\end{figure}

 First, we  derive the relations of $T_x$ for nodes of level $k$ and that for nodes of level $k+1$. 
For any subunit of level $k$ as shown in FIG.  \ref{sub_k}, its two boundary nodes (i.e., $A_k$ and $B_k$) are the only two nodes of level $k$, its nodes of level $k+1$ are boundary nodes of its $3$  subunits of level $k+1$(i.e., $A_k$, $B_k$, $O_k$ and $C_k$). Assuming $T_x$ for node  of level $k$(i.e., $T_{A_k}$, $T_{B_k}$) are known, we will analyze $T_x$ for node $x$ of level $k+1$(i.e., $O_k$ and $C_k$).

For any $k\geq 0$,  it is easy to obtain the following equation due to  Eqs. (\ref{MTT}),  (\ref{SO}), (\ref{SC}), (\ref{WO}) and(\ref{WC}).
\begin{eqnarray} \label{TOk}
T_{O_{k}}&=&S_{O_{k}}-\Sigma+N_t W_{O_{k}}  \nonumber  \\
&=&\frac{1}{2}(S_{A_{k}}+S_{B_{k}})-2\cdot6^{t-k-1}-\Sigma \nonumber  \\
&&+N_t\cdot \left[\frac{1}{2}(W_{A_{k}}+W_{B_{k}})-\frac{2^{t-k}}{3^{k+1}}\right] \nonumber  \\
&=&\frac{1}{2}(T_{A_k}+T_{B_k})-4\cdot 6^{t-k-1}-\frac{2^{t-k}}{3^{k+1}}
\end{eqnarray}
and
\begin{eqnarray} \label{TCk}
\hspace{-5mm}T_{C_{k}}&=&S_{C_{k}}-\Sigma+N_t W_{C_{k}} \nonumber  \\
\hspace{-5mm}&=&\frac{1}{2}(S_{A_{k}}+S_{B_{k}})+(3^t-3^{t-k})\cdot2^{t-k-1}-\Sigma\nonumber  \\
\hspace{-5mm}&&+N_t\cdot \left\{\frac{1}{2}(W_{A_{k}}+W_{B_{k}})+(1-\frac{1}{3^{k}})\cdot 2^{t-k-1}\right\} \nonumber  \\
\hspace{-5mm}\!&=&\!\frac{1}{2}(T_{A_k}\!+\!T_{B_k})\!+\!(3^t\!-\!3^{t-k})2^{t-k}\!+\!(1\!-\!\frac{1}{3^{k}})2^{t-k-1}
\end{eqnarray}

Note that $T_{A_0}=T_{B_0}$ and let $k=0$ in Eqs.(\ref{TOk}) and (\ref{TCk}), we find
\begin{eqnarray} \label{TCom0}
   T_{O_{0}}< T_{A_0}=T_{B_0}=T_{C_{0}}
\end{eqnarray}
As proved in Appendix \ref{Pro_ComTk}, we  find Eq.(\ref{ComTk}) holds for $k\geq1$.
\begin{equation}
      min\{T_{A_k},   T_{B_k}\}< T_{O_{k}}< max\{T_{A_k},   T_{B_k}\}\leq T_{C_k}
       \label{ComTk}
\end{equation}
Therefore, for $k\geq1$
   \begin{equation}
      min\{T_{A_k},   T_{B_k}\}=  min\{T_{A_k},   T_{B_k}, T_{O_{k}}, T_{C_{k}}\}
       \label{minTk}
\end{equation}

Let $ \Omega_k$ denote the set for nodes of level $k$ and note that $A_k$ and $B_k$ are the only two nodes of level $k$ in $\Lambda_k$,   $\{{A_k},   {B_k}, {O_{k}}, {C_{k}}\}$ represents all nodes of level $k+1$  in $\Lambda_k$,    Eq.(\ref{minTk}) implies: for any $k\geq1$,
 \begin{equation}\label{Tmink}
       min \{T_x: x\in\Omega_k\}=  min \{T_x: x\in\Omega_{k+1}\}
 \end{equation}
But Eqs.(\ref{TCom0})  shows
     \begin{equation}\label{Tmin1}
       T_{O_{0}}=min\{T_x: x \in \Omega_1\}<min\{T_x: x \in \Omega_0\}
     \end{equation}
Thus
\begin{equation}\label{Tmin}
       T_{O_{0}}=min\{T_x: x \in \Omega_1\}=min\{T_x: x \in \Omega\}
 \end{equation}
  Eqs.(\ref{TCom0})and (\ref{ComTk}) also shows
 \begin{equation}\label{Tmaxk}
       T_{C_{k}}=max\{T_x: x \in \Omega_{k+1}\} \qquad  k\geq0
 \end{equation}

 As proved in Appendix \ref{Pro_ComTCk}, we also find that: in any subunit $\Lambda_k$,  Eq.(\ref{ComTCk}) holds for any $k\geq0$.
 \begin{equation}\label{ComTCk}
  T_{C_{k}}\geq T_{C_{k+1}}
  \end{equation}
 Therefore, in subunit  $\Lambda_0$ (i.e. $G(t)$)
\begin{eqnarray}
       T_{C_{0}}&=&max\{T_x: x \in \Omega_1\} \nonumber  \\
       &\geq& T_{C_{t}}=max\{T_x: x \in \Omega_t\}\nonumber
 \end{eqnarray}
 But $\Omega_t=\Omega$ and $C_0\in\Omega$, we obtain
  \begin{equation}\label{Tmax}
       T_{C_{0}}=max\{T_x: x \in \Omega_t\}=max\{T_x: x \in \Omega\}
 \end{equation}
 Note:$T_{A_{0}}=T_{B_{0}}=T_{C_{0}}$ and ${A_{0}}, {B_{0}}, {C_{0}}$ and $O_0$  are also labeled as $A, B, C$ and $O$ in FIG. \ref{structure}. Eq.(\ref{Tmax}) and (\ref{Tmin}) shows that node $O_0$, which is  the hub  node of $G(t)$,  is best trapping site, nodes $A_0$, $B_0$ and $C_0$ , which is  the three boundary  nodes of $G(t)$, are  worst trapping sites.

 Comparing $T_{C_{0}}, T_{O_0}$  shown in Eqs.(\ref{MTTABC}) and (\ref{MTTO}), while $t\rightarrow \infty$, we have
 \begin{eqnarray}
T_{C_{0}}\approx 6\cdot T_{O_0}
\end{eqnarray}
which means that the location of target node has big effect on the trapping efficiency.

\section{Comparison the diffusion efficiency among all the nodes of T-fractal}
\label{sec:Comparison and analysis}

In these section, using the MDT  as the measure of trapping efficiency, we compare the trapping efficiency (i.e., the MDT) among all the nodes of  T-fractal and find the best trapping sites(i.e. nodes which have the minimum MTT) and the  worst trapping sites(i.e. nodes which have the maximum MDT).

Similarity to the analysis of trapping efficiency, we first derive the relations of $D_x$ for nodes of level $k$ and that for nodes of level $k+1$, and then compare $D_x$ for nodes of adjacent level.

Considering any subunit  of level $k (k\geq 0)$, which is  shown in FIG. \ref{sub_k},   it is easy to obtain the following equation due to  Eqs. (\ref{MDT}),(\ref{SO}), (\ref{SC}), (\ref{WO}) and (\ref{WC}).
\begin{eqnarray} \label{DOk}
 D_{O_{k}}&=&S_{O_{k}}+\Sigma-N_t W_{O_{k}}  \nonumber  \\
 &=&\frac{1}{2}(S_{A_{k}}+S_{B_{k}})-2\cdot6^{t-k-1}+\Sigma \nonumber  \\
 &&-N_t\cdot \left[\frac{1}{2}(W_{A_{k}}+W_{B_{k}})-\frac{2^{t-k}}{3^{k+1}}\right] \nonumber  \\
 &=&\frac{1}{2}(D_{A_{k}}+D_{B_{k}})+ {2^{t-k}}{3^{-k-1}}
\end{eqnarray}
and
\begin{eqnarray} \label{DCk}
 D_{C_{k}}&=&S_{C_{k}}+\Sigma-N_t W_{C_{k}} \nonumber  \\
 &=&\frac{1}{2}(S_{A_{k}}+S_{B_{k}})+(3^t-3^{t-k})\cdot2^{t-k-1}+\Sigma \nonumber  \\
 &&-N_t\cdot \left\{\frac{1}{2}(W_{A_{k}}+W_{B_{k}})+(1-{3^{-k}})\cdot 2^{t-k-1}\right\}\nonumber  \\
 &=&\frac{1}{2}(D_{A_k}+D_{B_k}) -(1-{3^{-k}})\cdot 2^{t-k-1}
\end{eqnarray}

Note that $D_{A_0}=D_{B_0}$ and let $k=0$ in Eqs.(\ref{DOk}) and (\ref{DCk}), we find
\begin{eqnarray} \label{DCom0}
   D_{O_{0}}> D_{A_0}=D_{B_0}=D_{C_{0}}
\end{eqnarray}
 As proved in Appendix \ref{Pro_ComDk}, we  find Eq.(\ref{ComDk}) holds for $k\geq1$.
  \begin{equation}
     D_{O_{k}}\!\leq\! min\{D_{A_k},   D_{B_k}\}\!<\!D_{O_{k}}\!<\!max\{D_{A_k},   D_{B_k}\}
       \label{ComDk}
   \end{equation}
Therefore, for $k\geq1$, we have
  \begin{equation}
     max\{D_{A_k},   D_{B_k}\}=max\{D_{A_k},  D_{B_k}, D_{O_{k}}, D_{C_{k}}\}
       \label{maxDk}
   \end{equation}
Because $A_k$ and $B_k$ are the only two nodes of level $k$ in $\Lambda_k$ and  $\{{A_k},  {B_k}, {O_{k}}, {C_{k}}\}$ represents all nodes of level $k+1$  in $\Lambda_k$,
Eqs.(\ref{maxDk}) implies: for any $k\geq1$
 \begin{equation}\label{Dmaxk}
       max \{D_x: x\in\Omega_k\}=  max \{D_x: x\in\Omega_{k+1}\}
 \end{equation}
But Eqs.(\ref{DCom0})  leads to
     \begin{equation}\label{Dmin1}
       D_{O_{0}}=max\{D_x: x \in \Omega_1\}>min\{D_x: x \in \Omega_0\}
     \end{equation}
Thus
\begin{equation}\label{Dmax}
       D_{O_{0}}=max\{D_x: x \in \Omega_1\}=max\{D_x: x \in \Omega\}
 \end{equation}
 Eqs.(\ref{DCom0})and (\ref{ComDk}) also shows
 \begin{equation}\label{Tmaxk}
       D_{C_{k}}=min\{T_x: x \in \Omega_{k+1}\} \qquad  k\geq0
 \end{equation}

 As proved in Appendix \ref{Pro_ComDCk}, we also find that: in any subunit $\Lambda_k$,  Eq.(\ref{ComDCk}) holds for any $k\geq0$.
 \begin{equation}\label{ComDCk}
  D_{C_{k}}\leq D_{C_{k+1}}
  \end{equation}
 Therefore
\begin{equation}\label{Dmin}
       D_{C_{0}}=min\{T_x: x \in \Omega\}
 \end{equation}
 Note: $D_{A_{0}}=D_{B_{0}}=D_{C_{0}}$ and  ${A_{0}}, {B_{0}}, {C_{0}}$ and $O_0$  are also labeled as $A, B, C$ and $O$ in FIG. \ref{structure}. Eq.(\ref{Dmax}) and (\ref{Dmin}) shows that node $O_0$  is worst diffusion site, nodes $A_0$, $B_0$ and $C_0$  are  best diffusion sites.

Comparing $D_{C_0}, D_{O_0}$  shown in Eqs.  (\ref{MDTABC}) and (\ref{MDTO}), while $t\rightarrow \infty$, we have
  \begin{eqnarray}
D_{A_0} \approx D_{O_0}
\end{eqnarray}
which means that the location of source node almost has no effect on diffusion efficiency.

\section{Conclusion}
\label{sec:4}

In this paper,we study unbiased discrete random walks on the T-fractal, Our effort is focused on the MFPT. We
 present new methods to  calculate the MFPT for any pair of nodes, the mean trapping time(MTT) for any target node and the mean diffusing time(MDT)  for any source node.
Using the MTT and the MDT as the measures of trapping efficiency and diffusion efficiency respectively, we compare the trapping efficiency and diffusion efficiency among  all nodes  of T-fractal and find the best (or worst)  trapping sites  and the best ( or worst) diffusing sites  which have the minimum MDT (or maximum MDT). Our results show that: the hub node of T-fractal is the best trapping site, but it is also the worst diffusing site, the three boundary  nodes  are the worst trapping sites, but they are also the best diffusing sites.  The methods  we present can also be used to solve the problems of MFPT on other self-similar trees.

\begin{acknowledgments}
The authors are grateful to the anonymous referees for their valuable comments and suggestions. This work was supported  by
the scientific research program of Guangzhou municipal colleges and universities under Grant No. 2012A022.
\end{acknowledgments}

\appendix
\section{Proof of Lemma \ref{Rec_Sk}  }
\label{Pof_Rec_sk}
\begin{figure}
\begin{center}
\includegraphics[scale=0.5]{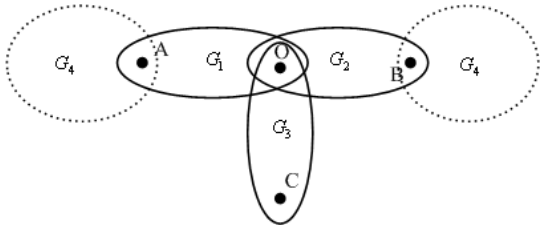}
\caption{Construction for subunit $\Lambda_{k-1}$: it is  composed of three  subunits of level $k$ labeled as  $G_1$,$G_2$,$G_3$ (oval with solid line in the figure), The subunit labeled as $G_4$ denote the the rest part of the T-fractal $G(t)$ except for the subunit $\Lambda_{k-1}$ (oval with dotted line in the figure).}
\label{Relation}
\end{center}
\end{figure}

Considering any subunit of level $k-1$ as shown in FIG. \ref{Relation}, it is  composed of three  subunits of level $k$. It is also connected with other part of the  T-fractal by its two boundary nodes (i.e. A and B in FIG. \ref{Relation}). In  this subunit, the two boundary nodes are the only two nodes of level $k-1$,  its nodes of level $k$ are boundary nodes of its $3$  subunits of level $k$(i.e., $A$, $B$, $O$ and $C$ in FIG. \ref{Relation}). Assuming $S_x$ for node  of level $k-1$(i.e., $S_A$, $S_B$) is known, we will analyze $S_x$ for node $x$ of level $k$(i.e., $O$ and $C$).

For simplify,  let denote
\begin{equation}
S_x(i)=\sum_{y \in G_i}{L_{yx}} \qquad \qquad i=1, 2, 3, 4
\end{equation}
Thus
\begin{equation}
S_x=S_x(1)+S_x(2)+S_x(3)+S_x(4)-2L_{xO} \label{SX}
\end{equation}

Because $A$ and $O$ are the two boundary nodes of subunit $G_1$, it is easy to know $S_O(1)=S_A(1)$. Note that the total numbers of nodes of  $G_1$ is $N_{t-k}$ and for any node $y\in G_1$,  $L_{yO}=L_{yB}-L_{OB}$. Therefore
\begin{eqnarray}
S_{O}(1)&=&\sum_{y \in G_1}{L_{yO}} \nonumber \\
&=&\sum_{y \in G_1}(L_{yB}-L_{OB}) \nonumber \\
&=&S_B-N_{t-k}L_{OB}
\label{SO1}
\end{eqnarray}
Thus
\begin{eqnarray}
S_O(1)&=&\frac{1}{2}[S_A(1)+S_B(1)-N_{t-k}L_{BO}]  \label{O1}
\end{eqnarray}
By symmetry, we have
\begin{eqnarray}
S_O(2)&=&\frac{1}{2}[S_A(2)+S_B(2)-N_{t-k}L_{AO}]  \label{O2}
\end{eqnarray}
For any  node $y\in G_3$,  $L_{yO}=L_{yB}-L_{OB}$ and $L_{yO}=L_{yA}-L_{OA}$. Note that $L_{OA}=L_{OB}$, we have
  $$L_{yO}=\frac{1}{2}(L_{yA}+L_{yB})-L_{OA}$$
Therefore
\begin{eqnarray}
S_{O}(3)&=&\sum_{y \in G_3}{L_{yO}} \nonumber \\
&=&\sum_{y \in G_3}[\frac{1}{2}(L_{yA}+L_{yB})-L_{OA}] \nonumber \\
&=&\frac{1}{2}[S_A(3)+S_B(3)]-N_{t-k}L_{AO}
\label{SO3}
\end{eqnarray}
For   any  node $y\in G_4$,    $L_{yO}=\frac{1}{2}(L_{yA}+L_{yB})$. Therefore
\begin{eqnarray}
S_O(4)=\frac{1}{2}[S_A(4)+S_B(4)]  \label{O4}
\end{eqnarray}
Note that $\Lambda_k$ is a copy of $G(t-k)$, one gets $L_{AO}=L_{BO}=L_{CO}=2^{t-k}$ and $N_{t-k}=3^{t-k}+1$, thus
\begin{eqnarray}
S_O&=&S_O(1)+S_O(2)+S_O(3)+S_O(4) \nonumber \\
     &=&\frac{1}{2}[S_A(1)+S_A(2)+S_A(3)+S_A(4)-2L_{AO}] \nonumber \\
     & &+\frac{1}{2}[S_B(1)+S_B(2)+S_B(3)+S_B(4)-2L_{BO}] \nonumber \\
     &&-2(N_{t-k}-1)L_{BO}  \nonumber \\
     &=&\frac{1}{2}(S_A+S_B)-2\cdot6^{t-k} \label{SO}
\end{eqnarray}
Similarly
\begin{eqnarray}
S_C(1)&=&S_O(1)+N_{t-k}L_{OC}  \label{C1} \\
S_C(2)&=&S_O(2)+N_{t-k}L_{OC}  \label{C2} \\
S_C(3)&=&S_O(3)  \label{C3} \\
S_C(4)&=&S_O(4)+[N_{t}-3N_{t-k}+2]L_{OC}  \label{C4}
\end{eqnarray}
Thus
\begin{eqnarray}
S_C&=&S_C(1)+S_C(2)+S_C(3)+S_C(4)-2L_{OC} \nonumber \\
     &=&[S_O(1)\!+\!S_O(2)\!+\!S_O(3)\!+\!S_O(4)]\!+\![N_{t}\!-\!N_{t-k}]L_{OC} \nonumber \\
     &=&S_O+[N_{t}-N_{t-k}]L_{OC} \nonumber \\
     &=&\frac{1}{2}(S_A+S_B)+[N_{t}-3N_{t-k}+2]L_{OC}   \nonumber \\
     &=&\frac{1}{2}(S_A+S_B)+(3^t-3^{t-k+1})\cdot2^{t-k} \label{SC}
\end{eqnarray}

Eqs.(\ref{SO}), (\ref{SC}) can be rewritten as a linear system
\begin{equation}\label{SOC}
  \left( \begin{array}{c} S_{O}\\S_{C} \end{array} \right)\!=\!\!\left(\!                
  \begin{array}{cc}
   1/2 & 1/2 \\
   1/2 & 1/2
  \end{array}
\!\right)\!
 \!\left(\! \begin{array}{c}  S_{A}\\S_{B} \end{array} \!\right)\!
 +\!\left(\! \begin{array}{c} -2\cdot6^{t-k}\\(3^t-3^{t-k+1})\cdot2^{t-k} \end{array} \!\right)\!
\end{equation}
We label the two boundary  nodes of $\Lambda_k$ as $A_k, B_k$ and building  mappings between boundary  nodes of $\Lambda_{k-1}$ and boundary  nodes of $\Lambda_{k}$, which are shown in FIG. \ref{subn_k}.

For example, for $i_k=2$, we have the following equivalence relation.
$$A_{k-1}\equiv A, B_{k-1}\equiv B, A_k\equiv O, B_k\equiv C$$
Let $\mathcal{S}^{(k)}\equiv(S_{A(k)} \quad S_{B(k)})^T$,
 Eqs.(\ref{SOC}) can be rewritten as
\begin{equation}\label{SK2}
  \mathcal{S}^{(k)}
=\left(                 
  \begin{array}{cc}
   1/2 & 1/2 \\
   1/2 & 1/2
  \end{array}
\right)
\mathcal{S}^{(k-1)}
+\left( \begin{array}{c} -2\cdot6^{t-k}\\(3^t-3^{t-k+1})\cdot2^{t-k} \end{array} \right)
\end{equation}
Therefore,  Eq. (\ref{SK}) holds  for $i_k=2$.

Similarly, we can verify that Eq. (\ref{SK}) holds  for $i_k=0, 1$.

\section{Proof of Lemma \ref{Rec_Wk} }
\label{Pof_Rec_wk}

Considering any subunit of level $k-1$ as shown in FIG. \ref{Relation}, assuming $W_x$ for nodes   of level $k-1$ (i.e. $W_A$, $W_B$) is known, we will analyze $W_x$ for node $x$ of level $k$ (i.e. $W_O$, $W_C$).
Let
\begin{equation}
W_x(i)=\frac{1}{2E_t}\cdot \sum_{y \in  G_i}{(L_{xy}\cdot d_y)} \qquad \qquad i=1, 2, 3, 4
\end{equation}
 Note: the degree for nodes which is the intersection of these subgraph were counted respectively in every subgraph. For example the degree of node $O$ is $1$ (not $3$ ) in subgraph $G_1$, $G_2$, $G_3$.
Thus
\begin{equation}
W_x=W_x(1)+W_x(2)+W_x(3)+W_x(4) \label{WX}
\end{equation}
Similar to the analysis of $S_O$, we find
\begin{eqnarray}
W_O(1)&=&\frac{1}{2}[W_A(1)+W_B(1)-\frac{E_{t-k}}{E_t}L_{BO}]  \label{WO1} \\
W_O(2)&=&\frac{1}{2}[W_A(2)+W_B(2)-\frac{E_{t-k}}{E_t}L_{AO}]  \label{WO1} \\
W_O(3)&=&\frac{1}{2}[W_A(3)+W_B(3)]-\frac{E_{t-k}}{E_t}L_{CO}  \label{WO3}\\
W_O(4)&=&\frac{1}{2}[W_A(4)+W_B(4)]  \label{WO4}
\end{eqnarray}
Thus 
\begin{eqnarray}
W_O&=&W_O(1)+W_O(2)+W_O(3)+W_O(4) \nonumber \\
     &=&\frac{1}{2}[W_A(1)+W_A(2)+W_A(3)+W_A(4)] \nonumber \\
     & &+\frac{1}{2}[W_B(1)+W_B(2)+W_B(3)+W_B(4)]\nonumber \\
     &&-2\frac{E_{t-k}}{E_t}L_{AO}  \nonumber \\
     &=&\frac{1}{2}(W_A+W_B)-\frac{2^{t-k+1}}{3^k} \label{WO}
\end{eqnarray}
Similarly
\begin{eqnarray}
W_C(1)&=&W_O(1)+\frac{E_{t-k}}{E_t}L_{CO}  \label{WC1} \\
W_C(2)&=&W_O(2)+\frac{E_{t-k}}{E_t}L_{CO}  \label{WC2} \\
W_C(3)&=&W_O(3)   \label{WC3} \\
W_C(4)&=&W_O(4)+\frac{E_t-E_{t-k+1}}{E_t}L_{CO}  \label{WC4}
\end{eqnarray}
But $E_{t-k+1}=3E_{t-k}$ and one gets 
\begin{eqnarray}
W_C&=&W_C(1)+W_C(2)+W_C(3)+W_C(4) \nonumber \\
     &=&[W_O(1)+W_O(2)+W_O(3)+W_O(4)]\nonumber \\
     &&+\frac{E_t-E_{t-k}}{E_t}L_{CO} \nonumber \\
     &=&W_O+\frac{E_t-E_{t-k}}{E_t}L_{CO} \nonumber \\
     &=&\frac{1}{2}(W_A+W_B)+(1-\frac{1}{3^{k-1}})\cdot 2^{t-k} \label{WC}
\end{eqnarray}

Similar to Appendix \ref{Pof_Rec_sk}, if we label the two boundary nodes of $\Lambda_k$ as $A_k, B_k$  and let
$\mathcal{W}^{(k)}\equiv(W_{A(k)} \quad W_{B(k)})^T$,
we can verify Eq. (\ref{WK}) holds for $i_k=0, 1, 2$.

\section{Derivation of  $S_{A_0}$ and $W_{A_0}$}
\label{Der_SW0}

 $A_0$ is one of the two nodes of level $0$ (i.e., A, B in FIG. \ref{structure}), it is also one of the two boundary nodes of $G(t)$. In order to tell the difference of $S_{A_0}$ (and $W_{A_0}$ ) for the T-fractal of different generation $t$,  let $S_{A}^t$ and $W_{A}^t$ denote the $S_{A_0}$ and $W_{A_0}$  in T-fractal of generation $t$. It is easy to know $S_{A}^0=1$ and  $W_{A}^0=\frac{1}{2}$. For $t\geq1$, according  to the self-similar structure shown in FIG. \ref{structure}, $S_{A}^t$  satisfies the following recursion relation.
\begin{equation}
S_{A}^t=S_{A}^{t-1}\!+\![S_{A}^{t-1}\!+\!(N_{t-1}-1)L_{BO}]\!+\![S_{A}^{t-1}\!+\!(N_{t-1}-1)L_{CO}]  \nonumber
\end{equation}
For the right side of the equation, the first item represents the summation for shortest path length between node $A$ and nodes in the subunit $G_1(t)$, the second item represents the summation for shortest path length between node $A$ and nodes in the subunit $G_2(t)$, the third item represents the summation for shortest path length between node $A$ and nodes in the subunit $G_3(t)$. Note that $L_{BO}=L_{CO}=2^{t-1}$ and $N_{t-1}=3^{t-1}+1$, we have
\begin{eqnarray}
S_{A_0}=S_{A}^t &=&3S_{A}^{t-1}+2\cdot6^{t-1}  \nonumber \\
      &=&3^2S_{A}^{t-2}+6^{t-1}+2\cdot6^{t-1}  \nonumber \\
      &=&\cdots \nonumber \\
      &=&3^tS_{A}^{0}+2\cdot6^{t-1}[2^{t-1}+\cdots+2^{-1}+1]  \nonumber \\
      &=&4\cdot6^{t-1}+3^{t-1}\label{SA}
\end{eqnarray} \par

Similarity, we find that  $W_{A}^t$  satisfies the following recursion relation.
\begin{eqnarray}
W_{A}^t&=&\frac{1}{2E_t} \sum_{y \in G(t)}({L_{Ay}\cdot d_y}) \nonumber\\
     &=&\frac{1}{2E_t}\left\{ \sum_{y \in G_1(t)}{L_{Ay}  d_y}+\sum_{y \in G_2(t)}{L_{Ay}  d_y}\right. \nonumber\\
     &&+\left. \sum_{y \in G_3(t)}{L_{Ay}  d_y} \right\}\nonumber\\
     &=&W_A^{t-1}+\frac{2^t}{3}   \label{WAT}
\end{eqnarray}
Hence
\begin{eqnarray}
W_{A_0}=W_A^t &=&W_A^{t-1}+\frac{2^t}{3}  \nonumber \\
      &=&W_A^{t-2}+\frac{2^{t-1}}{3}+\frac{2^t}{3} \nonumber \\
      &=&\cdots \nonumber \\
      &=&W_A^{0}+\frac{2}{3}+\frac{2^2}{3}+\cdots+\frac{2^t}{3}   \nonumber \\
      &=&\frac{2^{t+1}}{3}-\frac{1}{6}\label{WA}
\end{eqnarray}

\section{Proof of Eq.(\ref{ComTk}) }
\label{Pro_ComTk}

For any $k\geq 1$, according the following mappings for nodes of $\Lambda_k$ and $\Lambda_{k+1}$
\begin{eqnarray} \label{mapping}
\left \{                 
  \begin{array}{ll}
   A_{k+1}\equiv A_k, B_{k+1}\equiv O_k  & i_{k+1}=0 \\
   A_{k+1}\equiv O_k, B_{k+1}\equiv B_k & i_{k+1}=1 \\
   A_{k+1}\equiv O_k, B_{k+1}\equiv C_k & i_{k+1}=2
  \end{array}
\right.
\end{eqnarray}
we have
\begin{eqnarray} \label{CaseofTB_A}
T_{B_{k+1}}\!-\!T_{A_{k+1}}
\!=\!
\!\left \{ \!                
  \begin{array}{ll}
   T_{O_k}-T_{A_k} & i_{k+1}=0 \\
   T_{B_k}-T_{O_k} & i_{k+1}=1 \\
   T_{C_k}-T_{O_k} & i_{k+1}=2
  \end{array}
\right.
\end{eqnarray}
 Replacing $T_{O_k}$  and $T_{C_k}$  with Eqs.(\ref{TOk}), (\ref{TCk})respectively, we have
 \begin{eqnarray}\label{TAK1_BK1}
&&T_{B_{k+1}}\!-\!T_{A_{k+1}} \nonumber \\
&\!=\!&
\!\left \{ \!                
  \begin{array}{ll}
   \frac{1}{2}(T_{B_k}\!-\!T_{A_k})\!-\!4\cdot 6^{t-k-1}\!-\!\frac{2^{t-k}}{3^{k+1}} & i_{k+1}=0 \\
   \frac{1}{2}(T_{B_k}\!-\!T_{A_k})\!+\!4\cdot 6^{t-k-1}\!+\!\frac{2^{t-k}}{3^{k+1}} & i_{k+1}=1 \\
   (3^t\!-\!3^{t-k-1})2^{t\!-\!k}\!+\!(1\!-\!\frac{1}{3^{k+1}})2^{t\!-\!k\!-\!1} & i_{k+1}=2
   \end{array}
\right.
\end{eqnarray}

 For any $k\geq1$, we find that
\begin{eqnarray}
  \hspace{-4mm}&& |T_{B_k}-T_{A_k}| \geq 4\cdot 6^{t-k}+{2^{t-k+1}}{3^{-k}} \label{LHAK_BK} \\
  \hspace{-4mm}&&|T_{B_k}-T_{A_k}|\leq (3^t-3^{t-k})2^{t-k+1}+(1-3^{-k})2^{t-k} \label{UHAK_BK}
\end{eqnarray}
The  Eqs.(\ref{LHAK_BK}) and (\ref{UHAK_BK}) are proved by mathematical induction as follows.\par
Note that $T_{A_0}=T_{B_0}$, let  $k=0$ in Eq. (\ref{TAK1_BK1}), we obtain
\begin{eqnarray}
T_{B_{1}}-T_{A_{1}}
&=&
\left \{                 
  \begin{array}{ll}
   -4\cdot 6^{t-1}-\frac{2^{t}}{3} & i_{1}=0 \\
   4\cdot 6^{t-1}+\frac{2^{t}}{3} & i_{1}=1 \\
   4\cdot 6^{t-1}+\frac{2^{t}}{3} & i_{1}=2
  \end{array}
\right.
\end{eqnarray}
Thus Eqs.(\ref{LHAK_BK}) and (\ref{UHAK_BK}) holds for $k=1$. Assuming that Eqs.(\ref{LHAK_BK}) and (\ref{UHAK_BK}) hold for some $k\geq 1$, we will prove that  Eqs.(\ref{LHAK_BK}) and (\ref{UHAK_BK}) also hold for $k+1$. 

According to Eq.(\ref{CaseofTB_A}), $T_{B_{k+1}}-T_{A_{k+1}}$ has $3$ cases due to the different value of $i_{k+1}$.
 It is easy to verify Eqs.(\ref{LHAK_BK}) and (\ref{UHAK_BK}) hold for  $i_{k+1}=2$ due to Eq.(\ref{TAK1_BK1}).\par
  For $i_{k+1}=0$, substituting  $T_{B_k}-T_{A_k}$ with right side of Eq.(\ref{LHAK_BK}), we obtain
\begin{eqnarray}
 &&\hspace{-0mm}|T_{B_{k+1}}-T_{A_{k+1}}|\nonumber \\
  &=&\left|\frac{1}{2}(T_{B_k}-T_{A_k})\!-\!4\cdot 6^{t-k-1}-{2^{t-k}}{3^{-k-1}}\right| \nonumber \\
  &\geq & \frac{1}{2} \left[4\cdot 6^{t-k}+{2^{t-k+1}}{3^{-k}}\right]\!-\!4\cdot 6^{t\!-\!k\!-\!1}\!-\!\frac{2^{t-k}}{3^{\!-\!k\!-\!1}}\nonumber \\
  &= & 8\cdot 6^{t-k-1}+2\cdot{2^{t-k}}{3^{-k-1}}\nonumber \\
  &> &4\cdot 6^{t-k-1}+{2^{t-k}}{3^{-k-1}} \label{LHAK1-BK1}
\end{eqnarray}
Substituting  $T_{A_k}-T_{B_k}$ with right side of Eq.(\ref{UHAK_BK})
, we have
\begin{eqnarray}
 &&|T_{B_{k+1}}\!-\!T_{A_{k+1}}|\nonumber \\
 &=&\left|\frac{1}{2}(T_{B_k}-T_{A_k})\!-\!4\cdot 6^{t-k-1}-{2^{t-k}}{3^{-k-1}}\right| \nonumber \\
     &\leq & \frac{1}{2}\! \left[\!(3^t-3^{t-k})2^{t-k+1}+(1-3^{-k})2^{t-k}\!] \right]\! \nonumber \\
     &&\!+\! 4\cdot 6^{t-k-1}+{2^{t-k}}{3^{-k-1}} \nonumber \\
      &= &(3^t-3^{t-k-1})2^{t-k}+(1-3^{-k-1})2^{t-k-1} \label{UHAK1-BK1}
\end{eqnarray}
Therefore, Eqs.(\ref{LHAK_BK}) and (\ref{UHAK_BK}) hold for $i_{k+1}=0$.

Similarity, we can prove they both hold for  $i_{k+1}= 1$. Therefore, we obtain  Eqs.(\ref{LHAK_BK})  and (\ref{UHAK_BK})  hold for all the $3$ cases of $T_{B_{k+1}}-T_{A_{k+1}}$  which led to they both hold for any $k\geq 1$.

We now come back to prove Eq.(\ref{ComTk}). Without loss of generality,
 assuming $T_{B_k}\geq T_{A_k}$. Similar to the proof of Eq(\ref{LHAK1-BK1}) and (\ref{UHAK1-BK1}), we obtain 
   \begin{eqnarray}
  T_{O_{k}}-T_{A_k}&= &\frac{1}{2}(T_{B_k}-T_{A_k})-4\cdot 6^{t-k-1}-\frac{2^{t-k}}{3^{k+1}}>0  \nonumber
\end{eqnarray}
\begin{eqnarray}
 T_{B_{k}}-T_{O_k}&= &\frac{1}{2}(T_{B_k}-T_{A_k})+4\cdot 6^{t-k-1}+\frac{2^{t-k}}{3^{k+1}}>0 \nonumber
\end{eqnarray}
and
\begin{eqnarray}
T_{B_{k}}-T_{C_{k}}&= &\frac{1}{2}(T_{B_k}-T_{A_k})-(3^t-3^{t-k})2^{t-k}\nonumber \\
 &&-(1-3^{-k})2^{t-k-1}\nonumber \\
 &\leq &\frac{1}{2}\cdot[(3^t-3^{t-k})2^{t-k+1}+(1-3^{-k})2^{t-k}] \nonumber \\
 & &-(3^t-3^{t-k})2^{t-k}-(1-3^{-k})2^{t-k-1}  \nonumber \\
 &= &0 \label{UHBK-CK}
\end{eqnarray}
Therefore,  Eq.(\ref{ComTk}) holds while $T_{B_k}\geq T_{A_k}$. By symmetry,  Eq.(\ref{ComTk}) holds while $T_{B_k}\leq T_{A_k}$.

\section{Proof of Eq.(\ref{ComTCk})}
 \label{Pro_ComTCk}

According to structure of $\Lambda_k$, the proof can be divided into  $3$ cases due to the different values of $i_{k+1}$.  Without loss of generality,  assuming $T_{B_k}\geq T_{A_k}$.

Case I: for $i_{k+1}=0$
\begin{eqnarray} \label{HCK+10}
 T_{C_{k+1}} &=&  \frac{1}{2}(T_{A_{k+1}}+T_{B_{k+1}})+(3^t-3^{t-k-1})2^{t-k-1}\nonumber \\
       &&+(1-3^{-k-1})2^{t-k-2}\nonumber \\
      &=&\frac{1}{2}(T_{A_{k}}+T_{O_{k}})+(3^t-3^{t-k-1})2^{t-k-1}\nonumber \\
      &&+(1-3^{-k-1})2^{t-k-2}  \nonumber \\
      &=&\frac{3}{4}T_{A_{k}}+\frac{1}{4}T_{B_{k}}-2\cdot 6^{t-k-1}-\frac{2^{t-k-1}}{3^{k+1}}\nonumber \\
      & &+(3^t-3^{t-k-1})2^{t-k-1}+(1-3^{-k-1})2^{t-k-2} \nonumber \\
      &=&\frac{3}{4}T_{A_{k}}\!+\!\frac{1}{4}T_{B_{k}}\!+\!(3^t\!-\!3^{t\!-\!k})2^{t\!-\!k\!-\!1}\!+\!(1-3^{-k})2^{t\!-\!k\!-\!2}\nonumber
 \end{eqnarray}
Therefore 
\begin{eqnarray}
 &&T_{C_{k+1}}-T_{C_{k}}\nonumber \\
 &=&\!-\!\frac{1}{4}(T_{B_k}\!-\!T_{A_k})\!+\!(3^t\!-\!3^{t\!-\!k})2^{t\!-\!k\!-\!1}\!+\!(1\!-\!3^{\!-\!k})2^{t\!-\!k\!-\!2}\nonumber \\
 & &-(3^t-3^{t-k})2^{t-k}-(1-3^{-k})2^{t-k-1} \nonumber \\
 &\leq &-\frac{1}{4}\cdot\left(4\cdot 6^{t-k}+\frac{2^{t-k+1}}{3^k}\right)-(3^t-3^{t-k})2^{t-k-1}\nonumber \\
 & &-(1-3^{-k})2^{t-k-2}\nonumber \\
  &\leq& 0 \label{HCK+1-CK0}
\end{eqnarray}
Case II: for $i_{k+1}=1$
\begin{eqnarray} \label{HCK+11}
 T_{C_{k+1}}
      &=&\frac{1}{2}(T_{B_{k}}+T_{O_{k}})+(3^t-3^{t-k-1})2^{t-k-1} \nonumber \\
      &&+(1-3^{-k-1})2^{t-k-2}  \nonumber \\
     &=&\frac{3}{4}T_{B_{k}}\!+\!\frac{1}{4}T_{A_{k}}\!+\!(3^t\!-\!3^{t\!-\!k})2^{t\!-\!k\!-\!1}\!+\!(1\!-\!3^{-k})2^{t\!-\!k\!-\!2}\nonumber
 \end{eqnarray}
Hence
\begin{eqnarray}
 &&T_{C_{k+1}}-T_{C_{k}}\nonumber \\
 &=&\frac{1}{4}(T_{B_k}\!-\!T_{A_k})\!+\!(3^t\!-\!3^{t\!-\!k})2^{t\!-\!k\!-\!1}\!+\!(1\!-\!3^{\!-\!k})2^{t\!-\!k\!-\!2}\nonumber \\
 & &-(3^t-3^{t-k})2^{t-k}-(1-3^{-k})2^{t-k-1} \nonumber \\
 &\leq &\frac{1}{4}\cdot\left[(3^t-3^{t-k})2^{t-k+1}+(1-3^{-k})2^{t-k}\right]\nonumber \\
 & &-(3^t-3^{t-k})2^{t-k-1}-(1-3^{-k})2^{t-k-2} \nonumber \\
 &= &0 \label{HCK+1-CK1}
\end{eqnarray}
Case III: for $i_{k+1}=2$, note that $A_{k+1}\equiv O_k, B_{k+1}\equiv C_k$.  By symmetry, we have
\begin{eqnarray} \label{RCK+12}
 T_{C_{k+1}}=T_{C_{k}}
\end{eqnarray}
Thus Eq.(\ref{ComTCk}) holds  for $i_{k+1}=0, 1, 2$ and  Eq.(\ref{ComTCk}) holds for any $k\geq 0$ while $T_{B_k}\geq T_{A_k}$. Similarity, we can prove  Eq.(\ref{ComTCk}) holds while $T_{B_k}\leq T_{A_k}$.

 \section{Proof of Eq.(\ref{ComDk}) }
\label{Pro_ComDk}

According the  mappings for nodes of $\Lambda_k$ and $\Lambda_{k+1}$ as shown in Eq.(\ref{mapping}), we have
\begin{eqnarray}
D_{A_{k\!+\!1}}\!-\!D_{B_{k\!+\!1}}&\!=\!&
\!\left \{\!                 
  \begin{array}{cc}
   D_{A_k}-D_{O_k} & i_{k+1}=0 \\
   D_{O_k}-D_{B_k} & i_{k+1}=1 \\
   D_{O_k}-D_{C_k} & i_{k+1}=2
  \end{array}
\right.  \\
&\!=\!& \label{AK1_BK1}
\!\left \{\!                 
  \begin{array}{ll}
   \frac{1}{2}(D_{A_k}\!-\!D_{B_k})\!-\!\frac{2^{t-k}}{3^{k+1}} & i_{k+1}=0 \\
   \frac{1}{2}(D_{A_k}\!-\!D_{B_k})\!+\!\frac{2^{t-k}}{3^{k+1}} & i_{k+1}=1 \\
   \frac{2^{t-k}}{3^{k+1}}\!+\!(1\!-\!\frac{1}{3^{k}}) 2^{t-k-1} & i_{k+1}=2
  \end{array}
\right.
\end{eqnarray}
For any $k\geq1$, we find
\begin{eqnarray}
  &&|D_{A_k}-D_{B_k}| \geq {2^{t-k+1}}{3^{-k}}\label{LAK_BK} \\
  &&|D_{A_k}-D_{B_k}| \leq (1-{3^{-k}})2^{t-k} \label{UAK_BK}
\end{eqnarray}
The  Eqs.(\ref{LAK_BK}) and (\ref{UAK_BK}) are proved by mathematical induction as follows.

Note that $D_{A_0}=D_{B_0}$, let  $k=0$ in Eq. (\ref{AK1_BK1}), we obtain
$$|D_{A_{1}}-D_{B_{1}}|=\frac{2^{t}}{3}$$
Thus Eqs.(\ref{LAK_BK})and (\ref{UAK_BK})  hold for $k=1$. Assuming that Eqs.(\ref{LAK_BK})and (\ref{UAK_BK}) hold for some $k\geq 1$,  we will prove that  Eqs.(\ref{LAK_BK})and (\ref{UAK_BK}) also hold for $k+1$.

According to Eq.(\ref{AK1_BK1}), $D_{A_{k+1}}-D_{B_{k+1}}$ has $3$ cases due to the different value of $i_{k+1}$
and it is easy to verify Eqs.(\ref{LAK_BK}) and (\ref{UAK_BK}) hold for  the case $i_{k+1}=2$.

  For $i_{k+1}=0$, substituting  $D_{A_k}-D_{B_k}$ with right side of Eq.(\ref{LAK_BK}), we obtain
  \begin{eqnarray}
 |D_{A_{k+1}}-D_{B_{k+1}}|&= &\left|\frac{1}{2}(D_{A_k}-D_{B_k})-{2^{t-k}}{3^{-k-1}}\right| \nonumber \\
                         &\geq & \frac{1}{2}\cdot {2^{t-k+1}}{3^{-k}}-{2^{t-k}}{3^{-k-1}}\nonumber \\
                         &>& {2^{t-k}}{3^{-k-1}} \label{LAK1-BK1}
\end{eqnarray}
Substituting  $D_{B_k}-D_{A_k}$ with right side of Eq.(\ref{UAK_BK}), we have
\begin{eqnarray}
 |D_{A_{k+1}}-D_{B_{k+1}}|&= &\left|\frac{1}{2}(D_{A_k}-D_{B_k})-{2^{t-k}}{3^{-k-1}}\right| \nonumber \\
 &\leq&(1-{3^{-k}})\cdot 2^{t-k-1}+{2^{t-k}}{3^{-k-1}}\nonumber \\
 &= &(1-{3^{-k-1}})2^{t-k-1} \label{UAK1-BK1}
\end{eqnarray}
Therefore, Eqs.(\ref{LAK_BK}) and (\ref{UAK_BK}) hold for $i_{k+1}=0$.

Similarity, we can prove they both hold for  $i_{k+1}= 1$. Therefore,  Eqs.(\ref{LAK_BK})  and (\ref{UAK_BK})  hold for all the $3$ cases of $D_{B_{k+1}}-D_{A_{k+1}}$  which led to they both hold for any $k\geq 1$.

We now come back to prove Eq.(\ref{ComDk}). Without loss of generality,
 assuming $D_{A_k}\geq D_{B_k}$. Similar to the derivation of Eq.(\ref{LAK1-BK1}) and (\ref{UAK1-BK1}), we have
\begin{eqnarray}
  D_{A_k}-D_{O_{k}}&= &\frac{1}{2}(D_{A_k}-D_{B_k})-{2^{t-k}}{3^{-k-1}}>0
\end{eqnarray}
\begin{eqnarray}
 D_{O_{k}}-D_{B_k}&= &\frac{1}{2}(D_{A_k}-D_{B_k})+{2^{t-k}}{3^{-k-1}} >0
\end{eqnarray}
and
\begin{eqnarray}
 D_{C_{k}}-D_{B_{k}}&= &\frac{1}{2}(D_{A_k}-D_{B_k})-(1-\frac{1}{3^{k}})\cdot 2^{t-k-1}\nonumber \\
 &\leq &\frac{1}{2}\cdot(1-\frac{1}{3^k})2^{t-k}--(1-\frac{1}{3^{k}})\cdot 2^{t-k-1}  \nonumber \\
 &= &0 \label{UCK-BK}
\end{eqnarray}
Thus, for any $k\geq 1$, Eq.(\ref{ComDk}) holds while $D_{A_k}\geq D_{B_k}$. By symmetry,  Eq.(\ref{ComDk}) also holds while $D_{A_k}\leq D_{B_k}$.

 \section{Proof of Eq.(\ref{ComDCk})}
 \label{Pro_ComDCk}

  Without loss of generality,  assuming $D_{A_k} \geq D_{B_k}$. The proof of Eq.(\ref{ComDCk}) is divided into  $3$ cases due to the different values of $i_{k+1}$.

Case I: for $i_{k+1}=0$
\begin{eqnarray} \label{RCK+10}
 D_{C_{k+1}} &=&  \frac{1}{2}(D_{A_{k+1}}+D_{B_{k+1}}) -(1-\frac{1}{3^{k+1}}) 2^{t-k-2} \nonumber \\
      &=&\frac{1}{2}(D_{A_{k}}+D_{O_{k}}) -(1-\frac{1}{3^{k+1}}) 2^{t-k-2}  \nonumber \\
      &=&\frac{3}{4}D_{A_{k}}+\frac{1}{4}D_{B_{k}}+\frac{2^{t-k-1}}{3^{k+1}} -(1-\frac{1}{3^{k+1}})2^{t-k-2} \nonumber \\
      &=&\frac{3}{4}D_{A_{k}}+\frac{1}{4}D_{B_{k}}+(\frac{1}{3^{k}}-1) 2^{t-k-2}
\end{eqnarray}
Thus
\begin{eqnarray}
 &&D_{C_{k+1}}-D_{C_{k}}\nonumber \\
 &= &\frac{1}{4}(D_{A_k}-D_{B_k})+(\frac{1}{3^{k}}-1) 2^{t-k-2}+(1-\frac{1}{3^{k}}) 2^{t-k-1}\nonumber \\
 &\geq &\frac{1}{4}\cdot\frac{2^{t-k+1}}{3^k}+(1-\frac{1}{3^{k}}) 2^{t-k-2}  \nonumber \\
 &\geq& 0 \label{CK+1-CK0}
\end{eqnarray}
Case II: for $i_{k+1}=1$
\begin{eqnarray} \label{RCK+11}
 D_{C_{k+1}} 
      &=&\frac{1}{2}(D_{O_{k}}+D_{B_{k}}) -(1-\frac{1}{3^{k+1}}) 2^{t-k-2}  \nonumber \\
      &=&\frac{1}{4}D_{A_{k}}+\frac{3}{4}D_{B_{k}}+\frac{2^{t-k-1}}{3^{k+1}} -(1-\frac{1}{3^{k+1}}) 2^{t-k-2} \nonumber \\
      &=&\frac{1}{4}D_{A_{k}}+\frac{3}{4}D_{B_{k}}+(\frac{1}{3^{k}}-1)2^{t-k-2}
\end{eqnarray}
Thus
\begin{eqnarray}
 &&D_{C_{k+1}}-D_{C_{k}}  \nonumber \\
 &= &-\frac{1}{4}(D_{A_k}-D_{B_k})+(\frac{1}{3^{k}}-1) 2^{t-k-2}+(1-\frac{1}{3^{k}})2^{t-k-1}\nonumber \\
 &\geq &-\frac{1}{4}(1-\frac{1}{3^k})2^{t-k}+(1-\frac{1}{3^{k}}) 2^{t-k-2}  \nonumber \\
 &= &0 \label{CK+1-CK1}
\end{eqnarray}
Case III: for $i_{k+1}=2$, note that $A_{k+1}\equiv O_k, B_{k+1}\equiv C_k$.  By symmetry, we have
\begin{eqnarray} \label{RCK+12}
 D_{C_{k+1}}=D_{C_{k}}
\end{eqnarray}
Thus Eq.(\ref{ComDCk}) 
 holds for all the three case of $i_{k+1}=0, 1, 2$ while $D_{A_k} \geq D_{B_k}$. By symmetry,  it also holds while $D_{A_k}\leq D_{B_k}$.

\nocite{*}

\end{document}